\documentclass[aps,pra,twocolumn,twoside,nofootinbib,superscriptaddress]
              {revtex4}

\usepackage{amsmath,amsthm,amssymb}
\usepackage{bbm}
\usepackage{hyperref}

\newcommand{\<}{\langle}
\renewcommand{\>}{\rangle}

\newcommand{\be}{\begin{equation}}
\newcommand{\ee}{\end{equation}}
\def\ba#1\ea{\begin{align}#1\end{align}} 

\newtheorem{theorem}{Theorem}
\newtheorem{lemma}[theorem]{Lemma}
\newtheorem{definition}[theorem]{Definition}

\newtheorem{corollary}[theorem]{Corollary}

\newcommand{\A}{{\mathcal A}}
\newcommand{\C}{{\mathbb C}}
\newcommand{\cC}{{\mathcal C}}
\newcommand{\D}{{\mathcal D}}

\newcommand{\F}{{\mathbb F}}
\newcommand{\N}{{\mathcal N}}
\newcommand{\M}{{\mathcal M}}
\newcommand{\GU}{{\mathrm{GU}}}
\newcommand{\PGU}{{\mathrm{PGU}}}
\newcommand{\nmub}{N_{\mathrm{MUB}}}
\newcommand{\nnmub}{N_{\mathrm{NMUB}}}
\newcommand{\nmols}{N_{\mathrm{MOLS}}}
\newcommand{\normalin}{\unlhd}
\newcommand{\tr}{\mathop{\mathrm{tr}}\nolimits}
\newcommand{\syl}{\mathop{\mathrm{Syl}}\nolimits}
\newcommand{\onemat}{{\mathbbm 1}}

\begin{document}

\title{The limitations of nice mutually unbiased bases}

\author{Michael Aschbacher}
\email[]{asch@its.caltech.edu}

\affiliation{Department of Mathematics,
             California Institute of Technology,
             Pasadena, CA 91125, USA}

\author{Andrew M. Childs}
\email[]{amchilds@caltech.edu}

\author{Pawel Wocjan}
\email[]{wocjan@cs.caltech.edu}

\affiliation{Institute for Quantum Information,
             California Institute of Technology,
             Pasadena, CA 91125, USA}

\date[]{8 December 2004}


\begin{abstract}
Mutually unbiased bases of a Hilbert space can be constructed by
partitioning a unitary error basis.  We consider this construction
when the unitary error basis is a nice error basis.  We show that the
number of resulting mutually unbiased bases can be at most one plus
the smallest prime power contained in the dimension, and therefore
that this construction cannot improve upon previous approaches.  We
prove this by establishing a correspondence between nice mutually
unbiased bases and abelian subgroups of the index group of a nice
error basis and then bounding the number of such subgroups.  This
bound also has implications for the construction of certain
combinatorial objects called nets.
\end{abstract}

\maketitle

\section{Introduction}

Two orthonormal bases ${\cal B}$ and ${\cal B}'$ of the Hilbert space
$\C^d$ are called {\em mutually unbiased} if and only if
\be
  |\<\phi|\psi\>|^2=1/d
\ee
for all $|\phi\> \in {\cal B}$ and all $|\psi\> \in {\cal B}'$.  Let
$\nmub(d)$ denote the maximum cardinality of any set containing
pairwise mutually unbiased bases (MUBs) of $\C^d$.  It is an open
question to determine $\nmub(d)$ for every $d$.

It is well known that $\nmub(d)$ cannot exceed $d+1$
\cite{DGS:75,KL:78,Hoggar:82,WF:89,BBRV:02}. There exist constructions
that attain this upper bound when $d$ is a prime \cite{Ivanovic:81},
and more generally, when $d$ is a prime power
\cite{WF:89,CCKS:97,Zauner:99,BBRV:02,KR:03}.  In other words, we have
\be
  \nmub(p^e)=p^e+1
\ee
for any prime $p$ and $e \ge 1$.

For non-prime power dimensions, the maximal number of mutually
unbiased bases $\nmub(d)$ is not known---even the smallest case,
$d=6$, is unresolved.  The first construction of mutually unbiased
bases in non-prime power dimensions appears in \cite{Zauner:99,KR:03}.
If $d=mn$, then we have
\be
  \nmub(d) \ge \min \{\nmub(m), \nmub(n)\}
\,.
\label{eq:product}
\ee
For arbitrary $d$, let $\pi(d)$ denote the set of prime factors of
$d$, and let $d_p$ denote the largest power of $p \in \pi(d)$ that
divides $d$.  Then
\be
  \nmub(d) \ge \min_{p \in \pi(d)} \nmub(d_p)
             = \min_{p \in \pi(d)} d_p + 1
             =: N(d)
\,.
\label{eq:backPrimePower}
\ee
We will refer to this construction as the {\em reduce to prime power
construction}. In particular, this result implies that $\nmub(d)\ge 3$
for any dimension $d$.  (Another proof of this fact can be found in
\cite{BBRV:02}.)

Based on (\ref{eq:backPrimePower}), one might suspect that $\nmub(d)$
is given by $N(d)$ for any dimension $d$.  But this is false;
a counterexample is provided by the construction in \cite{WB:02},
which yields more MUBs for certain dimensions than the reduce to prime
power construction.  It was shown that for all square dimensions
$d=s^2$, $\nmub(d) \ge \nmols(s)+2$, where $\nmols(s)$ is the maximal
number of mutually orthogonal Latin squares of size $s$.  When
$d=26^2$, for example, this shows $\nmub(26^2) \ge 6$, whereas
$N(26^2)=5$.  Note that this construction also has consequences for
non-square dimensions since we can use the decomposition
(\ref{eq:product}).

For prime power dimensions $d=p^e$, there are two types of
constructions that attain the upper bound $d+1$. The first is based on
exponential sums in finite fields and Galois rings \cite{KR:03}.  In
\cite{Archer:03} it was shown that a natural generalization of this
construction to arbitrary dimensions cannot yield more MUBs than the
reduce to prime power construction.

The second construction which attains the maximal number of MUBs in
prime power dimensions is based on finding maximal commuting subsets
of matrices of a unitary error basis \cite{BBRV:02}.  This idea can be
applied in any dimension, but it is not known how many MUBs can be
produced in this way when the dimension is not a prime power.

In this paper we concentrate on the second construction in the case in
which the unitary error basis is a nice error basis.  A nice error
basis is a special type of unitary error basis with an underlying
group structure.  We show that the maximal number of MUBs produced by
partitioning a nice error basis, $\nnmub(d)$, cannot exceed the number
$N(d)$ produced by the reduce to prime power construction.  This shows
that if we want to construct a large number of MUBs by partitioning a
unitary error basis, that basis should be {\em wicked} (i.e., not
equivalent to any nice error basis).

The remainder of the paper is organized as follows.  In Section
\ref{sec:mubfromueb}, we review the construction of mutually unbiased
bases from a partition of a unitary error basis, and in particular,
from a nice error basis.  We also establish a connection between nice
mutually unbiased bases and sets of trivially intersecting abelian
subgroups of the index group of a nice error basis.  In Section
\ref{sec:grouptheory}, we prove the main result by establishing a
bound on the size of such sets.  Then, in Section \ref{sec:achieve},
we discuss examples that show the upper bound of $N(d)$ on $\nnmub(d)$
is achieved.  In Section \ref{sec:stronger}, we give a stronger bound
for the particular case where the group is abelian and its structure
is known.  In Section \ref{sec:nets}, we point out that our results
also provide bounds on the sizes of nets constructed in a particular
way, and show that a complete set of nice MUBs corresponds to an
affine translation plane.  Finally, we conclude in Section
\ref{sec:discussion} with a discussion of the results and some open
problems.

\section{Nice mutually unbiased bases}
\label{sec:mubfromueb}

We will consider mutually unbiased bases constructed from certain
kinds of unitary error bases.  A {\em unitary error basis} ${\cal E}$
is a basis of the vector space of complex $d\times d$ matrices that is
orthogonal with respect to the trace inner product.  In other words, a
set of unitary matrices ${\cal E}:=\{U_1=\onemat,U_2,\ldots,U_{d^2}\}$
is a unitary error basis iff
\be
  \tr(U_k^\dag U_l) = d \, \delta_{k,l} \,,\quad k,l \in \{1,\ldots,d^2\}
\,.
\ee
Two constructions of unitary error bases are known: nice error bases,
a group-theoretic construction due to Knill \cite{Knill:96a}; and
shift-and-multiply bases, a combinatorial construction due to Werner
\cite{Werner:00}.  There exist nice error bases that are not
equivalent to any shift-and-multiply basis, as well as
shift-and-multiply bases that are wicked \cite{KR:03b}.

In this paper we are concerned primarily with nice error bases, which
are unitary error bases with an underlying group structure.  We will
use a definition that appears different from, but is equivalent to,
the one proposed by Knill (cf.\ \cite{KR:00}).  To give this
definition, we begin with some background material on projective
representations.

Let $\GU_d(\C)$ be the $d$-dimensional general unitary group over the
complex numbers, and let $P:\GU_d(\C) \to \PGU_d(\C)$ be the
projection onto the projective general unitary group $\PGU_d(\C) =
\GU_d(\C) / Z(\GU_d(\C))$, where $Z(\cdot)$ denotes the center.  A
$d$-dimensional {\em projective (unitary) representation} of a finite
group $G$ is a homomorphism $\rho:G \to \PGU_d(\C)$.  Given any such
map, one can choose a finite preimage $\hat G$ of $\rho(G)$ in
$\GU_d(\C)$ with $P(\hat G)=\rho(G)$.  The group $\hat G$ is of {\em
central type} if $|\rho(G)|=d^2$ and $\tr \hat g=0$ for each $\hat g
\in \hat G - Z(\GU_d(\C))$.  If $\rho$ is faithful and some (and hence
each) preimage $\hat G$ is of central type, then we say $\rho$ is of
{\em central type}.  Note that a finite subgroup $\hat G  \le
\GU_d(\C)$ with $|\hat G|/|Z(\hat G)|=d^2$ is of central type iff the
character $\chi$ of $\hat G$ on $\C^d$ is irreducible iff $\chi(\hat
g)=0$ for each $\hat g \in \hat G - Z(\hat G)$.

Nice error bases can be defined as follows:
\begin{definition}[Nice error basis]
Let $G$ be a group of order $d^2$ with identity element $1$.  A subset
$\N \subset \C^{d \times d}$ is a {\em nice error basis} if there
exists a projective representation $\rho:G \to \PGU_d(\C)$ of central
type such that $\N = \{U_g: g \in G\}$, with $P(U_g)=\rho(g)$ and
$U_1=1$.
\end{definition}
\noindent
The group $G$ is called the {\em index group} of the nice error basis
$\N$.  Notice that for each distinct $U_g,U_h \in \N$, $U_g^\dag U_h
\in U_{g^{-1}h} Z(\GU_d(\C))$, and hence is of trace $0$, so $\N$ is a
unitary error basis.

Unitary error bases can be used to produce mutually unbiased bases
using the following construction:
\begin{lemma}
\label{lemma:constr_MUB}
Let ${\cal C}=\cC_1\cup\ldots\cup\cC_n$ with $\cC_k\cap
\cC_l=\{\onemat\}$ for $k\neq l$ be a set of $n(d-1)+1$ unitary
matrices that are mutually orthogonal with respect to the trace inner
product. Furthermore, let each class $\cC_k$ of the partition of
${\cal C}$ contain $d$ commuting matrices $U_{k,t}$, $0\le t\le d-1$,
where $U_{k,0}:=\onemat$. For fixed $k$, let ${\cal B}_k$ contain the
common eigenvectors $|\psi_k^i\>$ of the matrices $U_{k,j}$. Then the
bases ${\cal B}_k$ form a set of $n$ mutually unbiased bases, i.e.,
\be
  |\<\psi_k^i|\psi_l^j\>|^2 = 1/d \quad \mbox{ for $k\ne l$.}
\label{eq:unbiased}
\ee
\end{lemma}
\noindent
For a proof of this result, see \cite{BBRV:02,Grassl:04}.  In
Section~\ref{sec:achieve} we give a shorter proof of condition
(\ref{eq:unbiased}) for the special case of $d+1$ nice error bases.

We address the question of how many mutually unbiased bases can be
constructed when the set $\cC$ in Lemma~\ref{lemma:constr_MUB} is a
subset of a nice error basis.  We call such bases {\em nice mutually
unbiased bases}.  The main result of this paper is the following:

\begin{theorem}[Limitations of nice MUBs]
\label{thm:mublimits}
Let $\N$ be a nice error basis of $\C^{d \times d}$ with index group
$G$. Then the maximal number $\nnmub(d)$ of mutually unbiased bases
that can be obtained by partitioning a subset $\cC$ of $\N$ according
to Lemma~\ref{lemma:constr_MUB} is at most
\be
  N(d) = \min_{p \in \pi(d)} d_p + 1
\,.
\ee
\end{theorem}

We prove Theorem~\ref{thm:mublimits} in the next section.  To do so,
we first establish a connection between nice error bases and trivially
intersecting abelian subgroups of the index group:
\begin{lemma}
\label{lem:nebsubgroups}
Let $G$ be the index group of a nice error basis $\N$ and let $\M$ be
a set of $d$ pairwise commuting members of $\N$.  Then $A=P(\M)$ is an
abelian subgroup of $G$.
\end{lemma}

\begin{proof}
Since the elements of $\M$ are mutually commuting, they can be
simultaneously diagonalized.  The trace orthogonality of a unitary
error basis implies that the diagonals of the elements of $\M$, when
written in their common eigenbasis, must be pairwise orthogonal as
vectors in $\C^d$ with the standard inner product.  Since there can be
at most $d$ orthogonal vectors in a $d$-dimensional space, $\M$ is a
maximal commuting subset of $\N$.  As $\M \subseteq \M' := \N \cap
\<\M\>$ and $M := \<\M\>$ is abelian, $\M = \M'$ by the maximality of
$\M$.  But since $P$ is a homomorphism, this shows that $A=P(\M)=P(M)$
is an abelian group.
\end{proof}

Given this connection, we can produce upper bounds on the number of
nice MUBs by proving upper bounds on the number of trivially
intersecting abelian subgroups of the index group.

\section{Abelian subgroups of the index group}
\label{sec:grouptheory}

In this section, we establish the main result of the paper
(Theorem~\ref{thm:mublimits}) by bounding the number of trivially
intersecting abelian subgroups of order $d$ of a group $G$ of order
$d^2$.  Throughout, we let $\A$ denote a set of such subgroups.

For any group $H$ and $p \in \pi(|H|)$, let $O_p(H)$ denote the
largest normal $p$-subgroup of $H$, and let $\syl_p(H)$ denote the set
of Sylow $p$-subgroups of $H$.  Also, let
\be
  E_p(H) := \{h \in H : h^p = 1\}
\label{eq:order1orp}
\ee
be the set of elements of $H$ of order 1 or $p$.

First we observe that $G$ can be written as the product of two of the
members of $\A$, and that a similar decomposition holds for certain
Sylow $p$-subgroups.

\begin{lemma}
\label{lem:decompose}
Consider $A,B \in \A$ with $A \ne B$.  Then $G=AB$ (and hence $G$ is
solvable).  Furthermore, $P_{A,B} := O_p(A) O_p(B) \in \syl_p(G)$.
\end{lemma}

\begin{proof}
We have
\be
  d^2 = |G| \ge |AB| = \frac{|A| |B|}{|A \cap B|} = d^2
\,,
\ee
so $AB=G$.  Since $G$ can be written as the product of abelian groups,
it is solvable (see for example \cite[13.3.2]{Sco64}).  Furthermore,
\cite[13.2.5]{Sco64} implies $P_{A,B} \in \syl_p(G)$.
\end{proof}

Now we construct a new group $G_p$ and a set of subgroups $\A_p$ that
will be easier to work with.

\begin{lemma}
\label{lem:reduce}
Suppose $|\A| \ge 2$.  For any $p \in \pi(d)$, let $\A_p := \{O_p(A) :
A \in \A\}$ and $G_p := \<\A_p\>$.
Then $|G_p|=d_p^2$, $\A_p$ is a set of abelian subgroups of $G_p$ of
order $d_p$ such that $|A_p \cap B_p|=1$ for all distinct $A_p,B_p \in
\A_p$, and the map $A \mapsto O_p(A)$ is a bijection of $\A$ with
$\A_p$ (so that in particular, $|\A|=|\A_p|$).
\end{lemma}

\begin{proof}
Let $A,B,C \in \A$ with $A \ne B$.  By Lemma~\ref{lem:decompose},
$P_{D,E}$ is a group for all distinct $D,E \in \{A,B,C\}$.  Thus
\ba
  P &:= P_{A,B} O_p(C) \\
    &= O_p(A) O_p(B) O_p(C) \\
    &= O_p(A) O_p(C) O_p(B) \\
    &= O_p(C) O_p(A) O_p(B) \\
    &= O_p(C) P_{A,B}
\,,
\ea
so $P$ is a group.  Since $|P|$ divides $|P_{A,B}||O_p(C)|$, $P$ is a
$p$-group.  Furthermore, since $P_{A,B} \in \syl_p(G)$, $P=P_{A,B}$.
Thus $G_p=\<\A_p\>=P_{A,B}$ for any distinct $A,B \in \A$, and the
lemma follows.
\end{proof}
\noindent

Now we give the bound for $p$-groups, which by Lemma~\ref{lem:reduce}
implies a bound for all groups.

\begin{lemma}
\label{lem:pgroupbound}
Let $G$ be a $p$-group for some prime $p$.  Then $|\A| \le \min_{A \in
\A} |E_p(A)|+1$.
\end{lemma}

\begin{proof}
The idea of the proof is to identify a subgroup $H \le G$ such that
partitioning the non-identity elements of $H$ according to membership
in $A \in \A$ bounds $|\A|$.  Let $X \le Z(G)$ with $|X|=p$, where
$Z(G)$ denotes the center of $G$ (such a subgroup must exist because
every $p$-group has a nontrivial center; see for example
\cite[5.16]{Aschbacher:00}).  For any fixed $A$, suppose $X \not\le A$
(we will show below that such an $X$ can always be chosen).  Then let
$H := E_p(AX)$.

To obtain the bound, we must compute $|H|$, $|H \cap A|$, and $|H \cap
D|$ for $D \in \D := \A - \{A\}$.  Note that $AX = A(AX \cap D)$ for
any $D \in \D$ (this follows because $A(AX \cap D) = AX \cap AD$ by
the modular property of groups, and $AD=G$ by
Lemma~\ref{lem:decompose}).  Furthermore, $AX \cap D$ has order $p$,
since $p|A| = |AX| = |A(AX \cap D)| = |A||AX \cap D|/|A \cap AX \cap
D| = |A||AX \cap D|$.  Therefore $|H \cap D|=|E_p(AX \cap D)|=|AX \cap
D|=p$.  Also, $H = E_p(A) (AX \cap D)$, and therefore $|H|=|E_p(A)|p$.
Finally, $H \cap A = E_p(A)$, so $|H \cap A| = |E_p(A)|$.  Since the
non-identity elements of the various $D \in \D$ are distinct, we have
\be
  |\D| \le \frac{|H|-|H \cap A|}{|H \cap D|-1}
       =   |E_p(A)|
\,,
\ee
which shows $|\A| = |\D|+1 \le |E_p(A)|+1$.

It remains to show that we can always choose $X$ such that $X \not\le
A$.  Supposing $X \le A$, we construct $Y \not\le A$ with $Y \le Z(G)$
and $|Y|=p$, and use $Y$ in place of $X$.  Let $C,D \in \D$ be
distinct, and let $Y:=CX \cap D$.  Since $X \not\le C$, we have $|Y| =
p$ by the same argument we used to show $|AX \cap D|=p$.  Since $|A
\cap D| = 1$, $Y \not\le A$.  Finally, $Y \le Z(G)$ since $y \in Y$
satisfies $y \in D$ and can also be written as $y=cx$ for $c \in C$
and $x \in X \le Z(G)$, so it commutes with any $c'd' \in CD = G$.
This completes the proof.
\end{proof}

Combining these results gives the following bound on the size of $\A$:

\begin{lemma}
\label{lem:snakebound}
$|\A| \le \min_{p \in \pi(d), A \in \A} |E_p(A)|+1$.
\end{lemma}

\begin{proof}
This follows directly from Lemmas~\ref{lem:reduce} and
\ref{lem:pgroupbound}.
\end{proof}
\noindent
Now we can easily derive our main result.  By
Lemma~\ref{lem:nebsubgroups}, a partition $\cC = \cC_1 \cup \cdots
\cup \cC_n$ of $\N$ as in Lemma~\ref{lemma:constr_MUB} corresponds to
the set $\A=\{A_i: 1\le i \le n\}$ of subgroups of $G$, where $A_i =
P(\cC_i)$.  Then since $|E_p(A)| \le d_p$ for any $A \in \A$,
Lemma~\ref{lem:snakebound} implies $n=|\A| \le d_p+1$ as desired.

\section{Achieving the bound}
\label{sec:achieve}
 
In this section we construct examples which show the upper bound of
$N(d)$ on $\nnmub(d)$ is achieved, proving that the bound is best
possible.

First, consider the case where $|\A|=d+1$, i.e., there is a complete
set of nice MUBs.  In this case, $G$ must be an elementary abelian
group, $G = Z_p \times \cdots \times Z_p$ for some prime $p$.

\begin{corollary}
\label{cor:elabelian}
Suppose $|\A|=d+1$.  Then $G$ is elementary abelian.
\end{corollary}

\begin{proof}
If $|\A|=d+1$, then Lemma~\ref{lem:snakebound} implies
$|E_p(A)|=d$ for each $A \in \A$.  Thus every element of each $A \in
\A$ has order $1$ or $p$.  But since $|\A|=d+1$ and the distinct
members of $\A$ intersect trivially, every element of $G$ must appear
in some $A \in \A$.  Therefore every element of $G$ has order $1$ or
$p$.

Now let $X \le Z(G)$ with $|X|=p$, and choose $A \in \A$ with $X \ne
A$.  Arguing as in the proof of Lemma~\ref{lem:pgroupbound}, $AX-A$ is
partitioned by the subgroups $AX \cap D$ for $D \in \A - \{A\}$.  As
$X \le Z(G)$, $A$ centralizes $X \cap D$, so $X \cap D$ is in the
center of $AD=G$.  Then $A \le \<X \cap D: D \in \A-\{A\}\> \le Z(G)$,
so $G=AD$ is abelian, and in particular, elementary abelian.
\end{proof}
 
Now we show that $\nnmub(d)=d+1$ when $d=p^e$ is a prime power.  In
this case we know that $G$ must be elementary abelian, and we want to
show that this group has a nice error basis that can be partitioned
according to Lemma~\ref{lemma:constr_MUB}.  Such a partition was
constructed in \cite{BBRV:02}. Here we give a nonconstructive
existence proof based on some well-known group-theoretic facts and
then a more concrete construction along the lines of \cite{GHW:04}.

Let $Q$ be an extraspecial $p$-group of order $p^{1+2e}$. Then it is
known that $Q$ has a faithful irreducible representation of dimension
$d=p^e$ (see \cite[34.9]{Aschbacher:00}).  The group $G:=Q/Z(Q)$ is an
elementary abelian group of order $d^2$. The irreducible
representation of $Q$ gives rise to a projective representation of $G$
of central type. We can regard $G$ as a $2e$-dimensional vector space
over $\F_p$. It is also well known (see \cite[23.10]{Aschbacher:00})
that there is a symplectic form $f:G\times G\rightarrow \F_p$ on the
$\F_p$-space $G$ such that for $A\le G$, the preimage of $A$ in $Q$ is
abelian iff $A$ is a totally isotropic subspace of the symplectic
space $G$.  (A subspace $B$ is called {\em totally isotropic} iff
$f(u,v)=0$ for all $u,v\in B$.)

We see that a set $\A$ of $d+1$ abelian subgroups of order $d$ of $G$
partitioning $G$ corresponds to a set $\tilde{\A}$ of $d$-dimensional
totally isotropic subspaces partitioning the symplectic space $G$.
Then by Lemma~\ref{lemma:constr_MUB}, $\A$, and hence also
$\tilde{\A}$, determines a nice error basis $\N$ and a set $S$ of
$d+1$ mutually unbiased bases of $\C^d$.

In fact, in our special case this can be seen without appeal to
Lemma~\ref{lemma:constr_MUB}.  Given distinct $\tilde A_1, \tilde A_2
\in \tilde\A$, pick preimages $\hat A_i$ of $\tilde A_i$ in $Q$ and
complements $A_i$ to $Z := Z(\GU_d(\C))$ in $\hat A_i Z$.  Since $A_1$
acts (by conjugation) on $\hat A_2$, it acts regularly on the the set
of $1$-dimensional subspaces determined by the basis $\mathcal B_2$ of
common eigenvectors of all $\hat a_2 \in \hat A_2$.  Then the argument
in the proof of Theorem~2.1 in \cite{BBRV:02} shows that
$|\<\phi|\psi\>|^2=1/d$ for all $|\phi\> \in \mathcal B_1$ and
$|\psi\> \in \mathcal B_2$.

One set $\tilde{\A}$ can be constructed explicitly as follows.  Let
$\F_d$ be the finite field of order $d=p^e$, and let $T$ denote the
trace map from $\F_d$ to $\F_p$ (recall that the trace map is defined
by $T(\eta):=\eta+\eta^p+\ldots +\eta^{p^{e-1}}$ for all
$\eta\in\F_d$).  We can make $G$ into a $2$-dimensional symplectic
space over $\F_d$ by defining a symplectic form $g : G \times G
\rightarrow \F_d$. We can choose $g$ so that $f(u,v)=T(g(u,v))$ for
$u,v\in G$, and let $\tilde{\A}$ be the set of $d+1$ one-dimensional
$\F_d$-subspaces of $G$ (note that $1$-dimensional subspaces are
always totally isotropic). For distinct $A,B\in\tilde{A}$, $A\cap
B=0$, and so because $g$ is $0$ on $A$, so is $f=T\circ g$.
 
We will refer to $(G,f)$ as an $\F_p$-structure and to $(G,g)$ as an
$\F_d$-structure. Let $G:=H\times H$, where $H:=\F_p^e$ is the direct
product of $e$ copies of $\F_p$.  To define the $\F_d$-structure we
will identify $G$ with $G_{\F_d}:=\F_d\times \F_d$ via a suitable map
$\phi$ specified below.

Define the generalized Pauli operators
\be
  X :=
   \sum_{k=0}^{p-1} |k\>\< k+1|\,, \quad Z := \sum_{k=0}^{p-1} \omega^{k}
   |k\>\< k|\,, 
\ee
where $\omega$ is a $p$th root of unity. Denote the elements of
$G$ by $(x,z):=(x_1,\ldots,x_e, z_1,\ldots,z_e)$. Define the
map
\be
  \rho(x,y) := X^{x_1} Z^{z_1}\otimes \cdots \otimes X^{x_e} Z^{z_e} \,.
\label{eq:tensorGenPauli}
\ee
Then the set $\N:=\{ \rho(x,z) : (x,z)\in G\}$ is a nice error basis
with index group $G$. We define the map $f : G \times G \to \F_p$ as
\be
  f\big((x,z),(x',z')\big)) := \sum_{i=1}^e x_i z'_i - x'_i z_i\,.
\ee
The group $G$ together with the symplectic form $f$ is a symplectic
space of dimension $2e$ over $\F_p$.  Using the fact $XZ=\omega ZX$ it
follows that two matrices $\rho(x,z)$ and $\rho(x',z')$ commute iff
$f((x,z),(x',z'))=0$.

To view $G$ as a $2$-dimensional symplectic space over $\F_d$ we need
to define a symplectic form $g: G_{\F_d}\times G_{\F_d}\rightarrow
\F_d$. Furthermore, $g$ should satisfy the condition
$f(u,v)=T(g(\phi(u),\phi(v))$ for all $u,v \in G$. To do this we need
some basic definitions \cite{LN}. Let $\{a_1,\ldots,a_e\}$ be a basis
of the extension field $\F_d$ over the prime field $\F_p$ and
$\{b_1,\ldots,b_e\}$ the dual basis, i.e.,
\be
\label{eq:dualbasis}
  T(a_i b_j) = \delta_{ij}
\,.
\ee
Define the map $\phi:G \to G_{\F_d}$ as
\be
  \phi(x_1,\ldots,x_e,z_1,\ldots, z_e):=
  (\alpha,\beta)
\,,
\ee
where $\alpha:=\sum_{i=1}^e x_i a_i$ and $\beta:=\sum_{i=1}^e z_i
b_i$.  The symplectic form $g : G_{\F_d} \times G_{\F_d} \rightarrow
\F_d$ can be now defined as
\be
  g((\alpha,\beta),(\alpha',\beta')):=\alpha\beta' - \alpha'\beta
\,.
\ee
Using the property (\ref{eq:dualbasis}) one can explicitly check that
$f((x,z),(x',z'))=T(g(\phi(x,z),\phi(x',z')))$.

A collection of $d+1$ one-dimensional subspaces of $G_{\F_d}$ is given
by the lines
\be
  L_{\Delta}:=\{(\alpha,\Delta \alpha) : \alpha\in\F_{d}\}
\ee
(the $d$ lines with slope $\Delta\in\F_{d}$) and
\be
  L_{\infty} := \{(0,\beta) : \beta\in \F_{d}\}
\ee
(the line with slope $\infty$). Due to the discussion above, the sets
\be
  \cC_{\Delta}:=\{ \rho(\phi^{-1}(\alpha,\beta)) : (\alpha,\beta) \in
    L_{\Delta}\} \,,\quad \Delta\in\F_d\cup\{\infty\}
\ee
form a partition of the nice error basis into $d+1$ trivially
intersecting sets containing $d$ commuting matrices each, and hence
specify a set of $N(d)$ nice MUBs of dimension $d=p^e$.
 
This construction also lets us achieve the upper bound of
Theorem~\ref{thm:mublimits} in the non-prime power case, using an idea
along the lines of the reduce to prime power construction. More
precisely, for any dimension $d$ there is an index group $G$ of order
$d^2$ with corresponding nice error basis $\N$ such that we can obtain
$N(d)$ nice MUBs by partitioning $\N$ according to
Lemma~\ref{lemma:constr_MUB}. This is seen as follows.
 
Let $G_i$ be the elementary abelian group of order $p_i^{2 e_i}$,
$\rho_i$ the map in (\ref{eq:tensorGenPauli}), and $\N_i$ the
corresponding nice error basis for $i\in\{1,\ldots,r\}$. Let $G:=G_1
\times\cdots\times G_r$, $\rho:=\rho_1\otimes\cdots\otimes\rho_r$, and
$\N:=\N_1\otimes\cdots\otimes\N_r$. Let ${\cal C}^{(i)}:=\{ {\cal
C}^{(i)}_1,\ldots,{\cal C}^{(i)}_{p_i^{e_i}+1}\}$ be a partition of
$\N_i$ into $p_i^{e_i}+1$ commuting subsets. Choose for each $i$ an
arbitrary subset ${\cal D}^{(i)}$ of ${\cal C}^{(i)}$ of size $N(d)$.  
Then the sets
\be
  {\cal D}_k:=\{ {\cal D}^{(i)}_k \otimes\cdots\otimes 
                 {\cal D}^{(i)}_k: 1 \le i \le r\}
\ee
for $k\in\{1,\ldots,N(d)\}$ are subsets of $\N$ satisfying the
conditions of Lemma~\ref{lemma:constr_MUB}.

\section{Stronger bound for abelian index groups}
\label{sec:stronger}

Although Theorem~\ref{thm:mublimits} is the best possible bound
depending only on $|G|$, improved bounds on the size of $\A$ can be
obtained when we know something about the structure of $G$.  Here we
produce an improved bound for the case of abelian index groups.
Define
\be
  \bar E_p(A) := \{a^{p^{e-1}}: a \in O_p(A)\}
\,.
\ee
Then we have
\begin{lemma}
\label{lem:normalsubgroups}
Let $G$ be a group of order $d^2$, and let $\A$ be a set of trivially
intersecting subgroups of $G$ of order $d$ with the additional
condition that $A \normalin G$ for each $A \in \A$.  Suppose $|\A|>2$.
Then $G=A \times B$ for all distinct $A,B \in \A$, all members of $\A$
are abelian and isomorphic, and $|\A| \le \min_{p \in \pi(d)}
|\bar E_p(A)|$ for $A \in \A$.
\end{lemma}

\begin{proof}
As in Lemma~\ref{lem:decompose}, $G=A_1 A_2$ for all distinct $A_1,A_2
\in \A$.  Then as $|A_1 \cap A_2|=1$ and $A_1,A_2 \normalin G$, $G =
A_1 \times A_2$.  Since $|\A| > 2$, there is $A_3 \in \A -
\{A_1,A_2\}$.  Let $\Pi_i:A_3 \to A_i$ (for $i \in \{1,2\}$) be the
projection of $A_3$ onto $A_i$ with respect to the decomposition
$G=A_1 \times A_2$.  As $|A_{3-i} \cap A_3|=1$, $\Pi_i$ is injective,
and as $|A_3|=|A_i|$, $\Pi_i$ is an isomorphism.  Thus all members of
$\A$ are isomorphic.  Furthermore, let $a \in A_1$ and $b \in A_3$.
Then $[a,\Pi_1(b)]=[a,b] := a^{-1} b^{-1} a b \in A_1 \cap A_3 =
\{1\}$ since $A_1,A_3 \normalin G$.  Since $\Pi_1$ is an isomorphism,
$A_1$ is abelian, and therefore all members of $\A$ are abelian.

By Lemma~\ref{lem:reduce}, we may assume without loss of generality
that $d$ is a power of $p$.  Let $p^e$ be the exponent of $A:=A_1$,
and choose $X \le A_2$ with $X \cong Z_{p^e}$.  Now we proceed as in
the proof of Lemma~\ref{lem:pgroupbound}, but with $H := \bar
E_p(AX)$.  We have $AX = A (AX \cap B)$ for all $B \in \A - \{A\}$, so
choosing a generator $b$ for $AX \cap B$, $\<b^{p^{e-1}}\>$ is of
order $p$ in $H$.  Furthermore, $H=\bar E_p(A) \<b^{p^{e-1}}\>$, so as
in the proof of Lemma~\ref{lem:pgroupbound}, $|\A| \le |\bar
E_p(A)|+1$.  Since all members of $\A$ are isomorphic, this bound
holds for any $A \in \A$, and the lemma follows.  \end{proof}

Using this lemma, we can give a bound on the number of mutually
unbiased bases constructed from any particular abelian index group.
Note that abelian groups must be of the form $G = H \times H$ to be
index groups of nice error bases \cite{KR:00}.  (In the case $|\A| >
2$, this also follows from Lemma~\ref{lem:normalsubgroups}.)

\begin{corollary}
\label{cor:abelian}
Let $G=H \times H$ with $H = Z_{d_1} \times \cdots \times Z_{d_k}$,
where $d_1,\ldots,d_k$ are prime powers (without loss of generality).
Let $\mu_p(H):=\max\{d_j: p | d_j\}$, and let $\nu_p(H):=|\{j: d_j =
\mu_p(H)\}|$.  Then $|\A| \le \min_{p \in \pi(d)} p^{\nu_p(H)} + 1$.
\end{corollary}

\begin{proof}
Since any subgroup of an abelian group is normal, we can apply
Lemma~\ref{lem:normalsubgroups}.  Noting that $|\bar E_p(A)| =
\nu_p(H)$, the result follows.
\end{proof}
\noindent

As a simple example of this corollary, consider the index group $Z_d
\times Z_d$, which has a nice error basis given by generalized Pauli
operators \cite{Knill:96a}.  Reference \cite{Grassl:04} showed that at
most three MUBs of dimension six can be produced by partitioning the
generalized Pauli operators with $d=6$.  More generally, the result
above shows that a nice error basis of $Z_d \times Z_d$ can be
partitioned to produce at most $\min_{p \in \pi(d)} p + 1$ mutually
unbiased bases.

\section{Implications for nets}
\label{sec:nets}

In this section we show that the group-theoretic arguments of
Section~\ref{sec:grouptheory} can also be used to give upper bounds on
the number of parallel classes of nets. A net is a combinatorial
object that has many similar properties to a set of MUBs.  Using this
similarity, it was shown in \cite{WB:02} how to construct MUBs from
nets.  Our results in this section give further connections between
MUBs and nets.  Specifically, we present bounds on the number of
parallel classes of nets constructed in a particular way, and we show
that a complete set of nice mutually unbiased bases corresponds to an
affine translation plane.
 
\begin{definition}[Net]
A {\em $(d,k;\lambda)$-net} is a set $X$ of $\lambda d^2$ points
together with a set ${\cal B}$ of subsets of $X$ {\em (blocks)} each
of size $\lambda d$. The set ${\cal B}$ is partitioned into $k$ {\em
parallel classes}, each containing $d$ disjoint blocks. Every two
non-parallel blocks intersect in exactly $\lambda$ points.
\end{definition}
 
The analogy between a net and a set of mutually unbiased bases is
clear.  A parallel class is analogous to an orthonormal basis in a
collection of MUBs, and the condition that the bases be unbiased
corresponds to the requirement that blocks from different parallel
classes intersect in the same number of points.
                                                                                
A net is also referred to as an {\em affine design}, where ``affine''
indicates that every two non-parallel blocks intersect in the same
number of points.  We will only consider nets with $\lambda=1$, which
we refer to as $(d,k)$-nets.
                                                                                
Our results give an upper bound on the maximal number of parallel
classes when we use the following construction with abelian subgroups:
                                                                                
\begin{lemma}
\label{lem:nicenetconstruction}
Let $G$ be a group of order $d^2$ together with a set $\A$ of
subgroups of $G$ of order $d$ such that distinct subgroups intersect
trivially.  Then the incidence structure whose points are the elements
of $G$ and whose blocks are the left cosets of the subgroups defines a
$(d,|\A|)$-net.
\end{lemma}
\noindent
We emphasize that whereas the nice MUB construction requires the
subgroups to be abelian, the construction of nets does not.
 
\begin{proof}
Let $A\in \A$. Clearly the left cosets $G/A$ form a parallel class
since the cosets are a partition of $G$.  Assume that $|\A|\ge 2$ and
let $A,B\in \A$ be any two distinct subgroups. These cosets can be
expressed as $b A$ and $a B$ for some $a\in A$ and $b\in B$ because
$G=AB=BA$. It remains to show that the left cosets $bA$ and $aB$
intersect in exactly one point, i.e., $|bA \cap aB|=1$.
 
Assume that $|bA \cap aB| \ne 1$.  Then there are distinct $a',a'' \in
A$ and distinct $b',b'' \in B$ such that $b a'= a b'$ and $b a'' = a
b''$. But this implies that $a' (b')^{-1}=a b^{-1}=a'' (b'')^{-1}$, so
that $a'=a''$ and $b'=b''$, which is a contradiction.  Therefore $|bA
\cap aB|=1$, which completes the proof.
\end{proof}
 
If we restrict our attention to sets $\A$ containing abelian
subgroups, then Lemma~\ref{lem:snakebound} shows that $(d,k)$-nets
constructed according to Lemma~\ref{lem:nicenetconstruction} must have
$k \le N(d)$.

A $(d,d+1)$-net is called an {\em affine plane}.  Constructions of
affine planes are known when $d$ is a prime power \cite{vLW:92}.  An
affine plane obtained from $d+1$ subgroups of a group $G$ according to
Lemma~\ref{lem:nicenetconstruction} is called an {\em affine
translation plane}. For $G$ abelian it is known that $G$ must be
elementary abelian for such subgroups to exist \cite{Andre:54}.  (Note
that this also follows from Corollary~\ref{cor:elabelian}.)  Thus a
maximal set of nice MUBs corresponds to an affine translation plane.
 
\section{Discussion}
\label{sec:discussion}

We have shown that partitioning a nice error basis cannot produce more
mutually unbiased bases than the reduce to prime power construction.
This result demonstrates that novel approaches (such as the
construction of \cite{WB:02}) are needed to improve upon the reduce to
prime power construction.

The problem of determining $\nmub(d)$ for $d$ not a prime power
remains wide open, and although we have ruled out further progress by
construction of nice MUBs, there are many alternatives.  One possible
avenue is to show how to extend a nice mutually unbiased basis by
adding more bases that do not come from the eigenvectors of operators
in the nice error basis.  However, no such extension is possible when
$d=6$ \cite{Grassl:04}, so it would be interesting to determine
whether nice MUBs can ever be extended.  Another possibility is to
find ways of partitioning wicked error bases.  This approach may be
promising as many wicked error bases exist \cite{KR:03b}.  Finally,
one could look for constructions of MUBs that are not directly based
on partitioning unitary error bases, as in \cite{WB:02}.

\acknowledgments

MA is supported by the National Science Foundation under Grant No.\
DMS-0203417, and AMC and PW are supported by the National Science
Foundation under Grant No.\ EIA-0086038.


\newcommand{\noopsort}[1]{} \newcommand{\printfirst}[2]{#1}
  \newcommand{\singleletter}[1]{#1} \newcommand{\switchargs}[2]{#2#1}

\end{document}